\documentclass[11pt]{revtex4}
\usepackage{bm,latexsym,amsmath,amssymb,amsfonts,mathrsfs}

\newcommand{\bi}{\bibitem}
\newcommand{\be}{\begin{eqnarray}}
\newcommand{\ee}{\end{eqnarray}}

\newcommand{\rv}{\rho_{vac}}
\newcommand{\mqphi}{m^2_\phi}
\newcommand{\mqpl}{m^2_{Pl}}
\newcommand{\gmunu}{g_{\mu\nu}}

\begin{document}

\title{Dynamical vacuum energy via adjustment mechanism}
\author{A.D. Dolgov and F.R. Urban}
\affiliation{ITEP, Bolshaya Cheremushkinskaya 25, 117218, Moscow, Russia\\
INFN Sezione di Ferrara, via Saragat 1, 44100 Ferrara, Italy\\
Dipartimento di Fisica, Universit\`a di Ferrara, via Saragat 1, 44100 Ferrara, Italy}

\begin{abstract}

A new mechanism of adjustment of vacuum energy down to the observed value from an 
initially huge one is considered. The mechanism is based on a very strong variation 
of the gravitational coupling constant in very early universe. The model predicts 
that the non--compensated remnant of vacuum energy changes very slowly at late 
stages of the cosmological evolution and is naturally close to the observed one. 
Asymptotically the effective vacuum energy tends to a negative value, so the
cosmological expansion should stop and turn into contraction. There remains the
problem of introduction of the usual matter into the model and therefore realising realistic
cosmology.

\end{abstract}

\maketitle

\section{Introduction}

The life--story of vacuum energy is quite dramatic.
After it was introduced in 1918 by Einstein~\cite{ein-lam} under the
name cosmological constant or Lambda--term, it did not live long. The
Hubble discovery of the cosmological expansion~\cite{hubble} made
Einstein to agree that the evolution of the universe is described
by the non--stationary Friedmann solution~\cite{friedmann} governed by
the energy density of the usual matter. After that and till almost the
end of the XX century vacuum energy was rejected by the ``establishment'',
though there were a few bright names such as Le Maittre, Eddington,
Bronshtein who believed in Lambda. A short renaissance of Lambda took
place in the beginning of 60s when an attempt was done to explain
an accumulation of quasars near red-shift $z \approx 2$ by non--zero
vacuum energy~\cite{qso-lam}. Later it was understood that the red--shift
distribution of quasars can be well explained with $\rv = 0$ and the
general attitude to non--zero vacuum energy became even worse than it
was before.

Renewed interest to vacuum energy came from quantum field theory.
Probably the first published paper on the subject was by
Zeldovich~\cite{zeld-lam}. He stressed that the energy of vacuum
quantum fluctuations results in infinitely large vacuum energy and
suggested that the problem could be solved by cancellation of bosonic
and fermionic contributions. This would be indeed true if the world
were strictly supersymmetric but, as we know, this is not the case.
Supersymmetry, if it was ever realised in nature, would have been broken at the scale $M_{SUSY} \geq 1$ TeV and vacuum
energy in broken state should be about
\be
\rv^{(SUSY)} \sim M^{4}_{SUSY} \geq 10^{12}\,\,{\rm GeV}^4 \, .
\label{rho-susy}
\ee
In the case of locally realised supersymmetry, i.e.\
supergravity (SUGRA), vacuum energy is allowed to be small but at the
expense of unbelievably accurate fine-tuning. The
natural value of vacuum energy in such theories is about
$\rv^{(SUGRA)} \sim M_{Pl}^4 \sim 10^{76}$ GeV$^4$. To fit the observational
data (see below) the fine tuning must be precise with the accuracy of
$10^{-123}$!

A very strong argument in favour of non--trivial vacuum properties
is presented by Quantum Chromodynamics (QCD). According to this
theory, which perfectly agrees with experiment, the vacuum is not
empty. It is filled by quark~\cite{quark} and gluon
condensates~\cite{gluon}. The energy density of these QCD condensates
is negative and is about
\be
|\rv^{(QCD)}|\sim \Lambda_{QCD}^4 \sim 10^{-3}\,\,{\rm GeV}^4 \, .
\label{rho-qcd}
\ee
Existence of such condensates is
an experimentally established fact and the greatest mystery is what
else ``lives'' in vacuum whose energy compensates
$\rv^{(QCD)}$ down to $10^{-44}$ of the QCD contribution. It cannot
be any field related to QCD because all light fields are observed in
experiment, while heavy ones simply cannot achieve compensation with
such a precision.

So such compensating field, $\Phi$, cannot have colour
interaction and most naturally it is coupled to vacuum energy--momentum
tensor, $T_{\mu\nu}^{(vac)}$,
through gravity as any field should. But the coupling must
be arranged in such a way that the non--zero curvature induced by
vacuum creates a condensate of $\Phi$ whose energy in turn kills
the source. This is the idea of adjustment mechanism suggested a
quarter of century ago~\cite{ad-cam}. Since then a few dozens of
different models of adjustment have been studied, for a review
see e.g.\ ref.~\cite{ad-rev}. One may also refer to~\cite{ed} for a broader review on different alternative mechanisms available in the literature. Unfortunately none of the models
led to realistic cosmology without an additional fine--tuning.
On the other hand, such models successfully compensated a preexisting
vacuum energy down to the terms of the order of the critical energy
density $\rho_c \sim M_{Pl}^2/t^2$. This is a general feature of
adjustment mechanism. In this sense the prediction about cosmological
dark energy was done long before its discovery by astronomers.

Observational data in favour of non--zero vacuum or, generally speaking,
dark energy were accumulated during the last 10 years. There are several
independent kinds of observations, which are all best explained if the
cosmological energy is dominated by an unknown form of the so called dark
energy, which may be simply vacuum energy or something similar to it.
One can distinguish between the two by the equation of state expressing
the pressure density, $p$, through the energy density, $\rho$:
\be
p = w \rho \, .
\label{eqn-of-state}
\ee
For vacuum energy $w=-1$ and $\rho$ remains constant in the
course of cosmological expansion. If $w \neq -1$ and, moreover,
$w = w(t)$, such unknown cosmological energy is called dark energy.
According to the data $w$ is compatible with (-1) with better than
10\% accuracy~\cite{wmap}. For an up--to--date and critical review of the current status of observations on the subject see~\cite{subir}.

Here we propose a new model of dynamical adjustment of
vacuum energy down to the observed value. The model is based on a scalar
field $\Phi$ which is non--minimally coupled to the curvature scalar,
$V(\Phi)R$, and in this sense it is close to the original paper~\cite{ad-cam}
but the model avoids many of the shortcomings of that mechanism. In particular,
the Weinberg no--go theorem~\cite{sw-rev} for adjustment models based
on a scalar field can be evaded. Moreover, cosmology based on the
suggested mechanism may be realistic.

\section{Lagrangian and equations of motion \label{s-eq-of-mot}}

The model of adjustment of vacuum energy down to (almost) zero,
which is considered here,
is based on a simple Lagrangian of a scalar field $\phi$ non--minimally
coupled to gravity:
\be
{\cal L} = \frac{1}{2} (\partial \phi)^2 - U(\phi) - V(\phi) R \, ,
\label{lagrange}
\ee
where $R$ is the curvature scalar.

The equation of motion for field $\phi$ has the form
\be
D^2 \phi + U'(\phi) + RV'(\phi) = 0 \, ,
\label{D2-phi}
\ee
where prime means derivative with respect to $\phi$ and
$D$ is the covariant derivative in the gravitational field. The latter
is governed by the equation:
\be
2 V(\phi)\left(R_{\mu\nu} - \frac{1}{2} g_{\mu\nu} R\right) =
\partial_\mu\phi\, \partial_\nu\phi -
 \frac{1}{2} \gmunu
\left[ \partial_\alpha\phi\, \partial^\alpha\phi - 2U(\phi)\right] \nonumber\\
- 2 \left(\gmunu D^2 - D_\mu D_\nu\right) V(\phi) + T_{\mu\nu} \, .
\label{ein-eq}
\ee
where $T_{\mu\nu}$ is the energy--momentum tensor of matter defined by
\be
T_{\mu\nu} = - \frac{2}{\sqrt g} \frac{\delta S_m}{\delta g^{\mu\nu}} \, .
\ee
The Minkowski metric is taken to be $\eta_{\mu\nu} = {\rm diag}(1,-1,-1,-1)$, 
and $g = - \det g_{\mu\nu}$.

To avoid misundersanding we would like to note that the vacuum energy term 
$\rho_{vac} \, g_{\mu\nu}$ is usually inserted into $T_{\mu\nu}$, but it is evident
that it can be added as a constant to $U(\phi)$ as well.

In what follows we will consider a homogeneous case described by the
Friedmann--Robertson--Walker (FRW) metric; the field $\phi$ is assumed to be a
function of time only.

Taking the trace of eq.~(\ref{ein-eq}) and using eq.~(\ref{D2-phi})
we find for the curvature scalar:
\be
R = \frac{\dot \phi^2 \left( 1 + 6 V''\right) - 4 U -
6 V' \,U' - T^\mu_\mu }
{2\left( V + 3 V'^2 \right)} \, .
\label{r-of-phi}
\ee
Now the equation of motion for $\phi$ takes the form:
\be
\ddot \phi + 3H\dot\phi + \frac{\dot\phi^2 V'\,\left( 1 +6 V''\right)
+ 2 U'V - 4 U V' -V' T^\mu_\mu}{2\left(V + 3V'^2\right)} =0 \, .
\label{ddot-phi}
\ee
The Hubble parameter is equal to:
\be
H = - \frac{V' \dot\phi}{2V} + \left[ \left(\frac{V' \dot\phi}{2V}\right)^2
+ \frac{\dot \phi^2 + 2 U + 2\rho_m}{12 V}\right]^{1/2} \, ,
\label{H}
\ee
where $\rho_m$ is the energy density of matter, which we
disregard below for a while.

The energy and pressure densities of the system are given, in the perfect fluid approximation, by:
\be
\rho_\phi &=& \frac{1}{2}\dot\phi^2 + U - 6 H \dot\phi V' \, ,\label{rho-phi}\\
P_\phi &=& \frac{1}{2}\dot\phi^2 \left[ 1 + 4 V'' - 
\frac{2 V'^2 \left( 1+ 6 V'' \right)}{V + 3 V'^2} \right] - 
U \left[ \frac{V - V'^2}{V + 3 V'^2} \right] + 2 H \dot\phi V' - 
\frac{2 U' V' V}{V + 3 V'^2} \label{p-phi} \, .
\ee

One can see that in the absence of matter or for relativsitic matter, i.e.\ for $T_\mu^\mu = 0$,
the stationary point of equation (\ref{ddot-phi}) is a solution of the equation:
\be
U(\phi_s) = \frac{V(\phi_s)}{2 V'(\phi_s)}\, U'(\phi_s) \, .
\label{eq-stat}
\ee
One can easily check that for positive vacuum energy, see below eq. (\ref{rho-v}),
the second derivative of the effective potential in the stationary point 
$\phi=\phi_s$, eq. (\ref{eq-stat}), is positive, i.e.\ this point is stable. 
If $T_\mu^\mu \neq 0$, the position of the stationary point would be shifted 
according to:
\be
U(\phi_s) + \frac{T^\mu_\mu}{4}= \frac{V(\phi_s)}{2 V'(\phi_s)}\, U'(\phi_s) \, .
\label{eq-stat2}
\ee
Since by construction $T_\mu^\mu$ includes only normal matter which decrease in
the course of cosmological expansion as $1/a^3$, where $a$ is the cosmological
scale factor, the omission of the $T_\mu^\mu$--term does not significantly change 
the result.

If by some reason the ratio $V/V'$ is small then the value of the
effective vacuum energy at equilibrium $U(\phi_s)$ is also small.
We will see that this is indeed the case for a certain choice of the
potentials $U(\phi)$ and $V(\phi)$ if the original vacuum
energy, $\rv$, is large.

There is nothing new in Lagrangian (\ref{lagrange}).
Many models of adjustment
starting from ref.~\cite{ad-cam} used similar ${\cal L}$.
Moreover, we take the potential $U(\phi)$ in
the simplest possible form, namely just the potential of free massive field:
\be
U(\phi) = \rho_{vac} \pm m_\phi^2 \phi^2/2 \, .
\label{U-of-phi}
\ee
Here $\rv$ is the initial vacuum energy density. If $\rv >0$ then we have to choose 
negative sign in front of the mass term
in eq.~(\ref{U-of-phi}); for negative $\rv$ the sign should be positive. 
It may seem that negative mass term would lead to instability because
the potential $U(\phi)$
is not bounded from below. We will see that it is not so because there
is a stable equilibrium point at finite $\phi=\phi_s$.
As we show in what follows, the resulting vacuum energy at this point,
$\rv^{(eff)} = U(\phi_s)$
is non--vanishing and automatically small, if the initial $\rv$ is large.

The important less trivial input is the form of the coupling of $\phi$ to
curvature $R$. We take it in the form:
\be
V(\phi) = \mqpl\,\exp\left(\frac{\phi^4 - \phi_s^4}{\mu^4}\right) \, ,
\label{V-of-phi}
\ee
though other similar types of the potential $V(\phi)$ are possible.
Here $\mqpl$ is the value
of the effective Planck mass at the stationary point of the equations of motion,
$\phi =\phi_s$. As we shall see shortly, the universe has not yet
reached the state where
$\phi = \phi_s$. Hence the value of the Planck mass now, $M_{Pl}$
would be different from that at the stationary point,
\be
M_{Pl}^2 = \mqpl \exp[(\phi_0^4 - \phi_s^4)/\mu^4] \, ,
\label{M-Pl}
\ee
where $\phi_0$ is the value of $\phi$ today.

In what follows we assume that the original vacuum energy is positive,
\be
\rv = M_v^4 > 0 \, ,
\label{rho-v}
\ee
and that $M_v$ is of the order of $M_{Pl}$, while the other relevant mass
parameters $m_\phi$ and $\mu$ are both of the order of TeV. Such a choice
of numerical values is not obligatory but an important thing is that
$M_v \gg (m_\phi,\,\, \mu)$.

The value of $\phi$ at the stationary point is:
\be
\phi_s^2 \approx \frac{2M_v^4}{m_\phi^2} + \frac{\mu^4 m_\phi^4}{M_v^4} - \frac{\mu^8 m_\phi^6}{128M_v^{12}} \, .
\label{phi-s-of-m}
\ee
The value of the
effective vacuum energy in the stationary point would be
\be
U(\phi_s) \approx -\left(\frac{\mu^2 \mqphi}{4 M_v^2}\right)^2 \, .
\label{U-phi-s}
\ee
With the chosen above ``natural'' numerical values (Planck scale for the vacuum, TeV scale for the fields) the vacuum energy
at the stationary point would be $(- 10^{-52})$ GeV$^4$, which is not
catastrophically far from the observed today density of dark
energy, $\rho_{DE} \sim 10^{-47}$ GeV$^4$. More troubling may be the
sign difference. However, we will see in the following section
that the universe spends cosmologically large time in the state
where $U(\phi)$ is positive
and close to the observed value (with a reasonable
choice of masses $m_\phi$ and $\mu$). So we have not yet reached the
asymptotic stationary point and most probably will never reach it
because the cosmological expansion will turn into contraction
before $\phi$ reaches $\phi_s$.

Let us also stress that such transplanckian value for $\phi_s$ is not a problem, since its potential is always smaller that $M_v^4$, and only powers of $U/M_v^4$ enter in the effective non--renormalisable potential, see~\cite{linde-book}. Notice further that at very early times, when $\phi \ll \phi_c$ (see below), this statement is no longer true due to the smallness of $V$, and a more complete quantum gravity description of the system is needed.

\section{Solution of equations of motion \label{sol}}

Numerical solution of eqs.~(\ref{ddot-phi}) and (\ref{H})
is difficult because of the huge variation of relevant
quantities due to exponential form of $V(\phi)$. Fortunately
an approximate analytical solution can be found in the
interesting range of $\phi$  values.

One can easily check that when $\phi< \phi_s$ and far from it, it rather
quickly rises in direction to $\phi_s$. During this stage the universe
expanded by a huge factor and at this regime $\phi$ plays the role
of the inflaton. So inflation is automatically implemented into the
model. We can see this by writing down the equation for the acceleration parameter $\ddot a$:
\be\label{acceleration}
\frac{\ddot a}{a} = -H^2 - \frac{R}{6} \simeq \frac{U - \dot\phi^2}{6V} \, ,
\ee
where the last step refers to the early stage of evolution. When $\phi \ll \phi_s$ the acceleration is huge and positive. Inflationary solutions for similar setup were studied in~\cite{alessio}, and found to be in excellent agreement with observations. Alternatively, one can make a conformal coordinate transformation $\bar g_{\mu\nu} = (\mqpl / V) \, \gmunu$ to the Einstein frame. In such a coordinate system the new gravity lagrangian will be just the usual one ${\cal L}_E = \mqpl\bar R$, while the potential for $\phi$ becomes, in close analogy to \cite{alessio}, $\bar U = m_{Pl}^4 \, (U / V^2)$, that is, a hybrid--like potential capable of giving inflationary expansion.

Moreover, in ref.~\cite{alessio}, with a somewhat differ choice for the potentials, a dynamical hierarchy between the effective Planck mass and the electroweak scale was generated. In our case we are primarily interested in a solution to the vacuum energy puzzle, and the variation of the effective Planck mass $m_{Pl}^2 (\phi) = V(\phi)$ at this stage is extremely more drastic. It would be of great interest if it were possible to unify the two proposals, even though at the moment it looks very challenging.

Here we study in some detail the behavior of the solution when $\phi$
approaches $\phi_s$, still being sufficiently far from it so that
$U(\phi)$ remains positive. We make the expansion
\be
\phi = \phi_s + \phi_1 \, ,
\label{phi-1}
\ee
where $\phi_1 <0$ and sufficiently small,
\be
\phi_1^2 < \frac{\mu^4}{6 \phi_s^2}\,\,\,{\rm or}\,\,\,
\phi_1 > \phi_d \equiv - \frac{\mu^2}{\sqrt{6} \phi_s} \, .
\label{phi-1-2}
\ee
If this inequality is fulfilled, we may neglect the
terms containing higher powers of $\phi_1$ and keep only linear terms.
It is convenient to introduce the notation:
\be
V'(\phi)/V(\phi) = 4\phi^3/\mu^4 \equiv K(\phi) \, ,
\label{K-of-phi}
\ee
and $K\equiv K(\phi_s)$. For a small $|\phi_1|$ the potential $V(\phi)$
can be written as
\be
V(\phi) \approx \mqpl\,\exp (K\phi_1) \, ,
\label{V-of-phi1}
\ee
while quadratic and higher terms in the exponent can be neglected.

When $\phi$ is sufficiently far from $\phi_s$ the following conditions
are fulfilled:
\be
V'' \ll 1,\,\,\,{\rm and}\,\,\, (V')^2 \ll V \, ,
\label{V''small}
\ee
and the equations of motion are significantly simplified.
The boundary value of $\phi_1=\phi_c$, when $V''(\phi) = 1$ is reached,
is equal to:
\be
\phi_1 <\phi_c \equiv -\frac{2}{K}\,\ln (K m_{Pl}) \, .
\label{phi-c}
\ee
Conditions (\ref{phi-1-2}) and (\ref{phi-c}) are compatible if $|\phi_d| >|\phi_c|$.
This is true when
\be
m_{Pl} K \simeq \frac{m_{Pl} M_v^6}{\mu^4 m_\phi^3} \lesssim \exp\frac{M_v^4}{\mu^2 m_\phi^2} \, .
\label{bound}
\ee
These conditions simplify the solution of the equations of motion but they are
not necessary otherwise and the model may still operate even if they are not
fulfilled.

We shall remark on the following detail before studying the different regimes of interest. Notice that in order to expand further the exponential in (\ref{V-of-phi1}) as $1 + 4\phi_s^3 \phi_1 / \mu^4 + \dots$ we would need $|\phi_1| \ll \mu^4 / \phi_s^3$. Since, as it will be soon clear, the interesting cosmological epochs are close to $|\phi_c| \gg \mu^4 / \phi_s^3$ we are not allowed to do so. This is to say that the exponential factor plays a key r\^ole in our setup.

When $\phi_1$ approaches $\phi_c$ being still larger by the absolute value,
i.e.\ $|\phi_1|>|\phi_c|$, equation (\ref{ddot-phi}) takes the form:
\be
\ddot \phi_1 + \frac{\sqrt{3}}{2}\,\dot\phi_1 \left(\frac{\dot \phi_1^2
+ 2 U}{V}\right)^{1/2} +
U' + \frac{V'}{2V}\,\left( \dot \phi^2_1 - 4 U\right) = 0 \, .
\label{ddot-phi1}
\ee
Remember that $V\sim \exp (K\phi_1)$ is exponentially small.

Potential (vacuum--like) energy at this stage is positive and is equal to:
\be
U \approx -\frac{\mu^4 m_\phi^4}{16 M_v^4} - m^2_\phi \phi_s \phi_1 \, .
\label{U-of-phi-1}
\ee
In the equation of motion (\ref{ddot-phi1}) the first constant term in $U$
cancels with the $U'V/2V' $ and only the term proportional to $\phi_1$
remains. Moreover, when $\phi_1$ approaches to $\phi_c$ from below, the first
term in $U$, eq.(\ref{U-of-phi-1}), is small in comparison with the second one
and at $\phi_1 = \phi_c$ the potential is approximately:
\be
U(\phi_s+\phi_c) \approx \frac{\mu^4 m_\phi^4}{4 M_v^4}\,\ln K m_{Pl} \, .
\label{U-of-phi-c}
\ee

Such vacuum energy is even closer to the observe value thanks to the large value of $K$ in the logarithm. It can be easily tuned down to the present day cosmological constant e.g.\ by increasing $m_\phi$ by a factor of 3.

Since the Hubble parameter at this stage is very big, the equation of motion
can be solved in slow roll approximation:
\be
\ddot\phi &\ll& 3H\dot\phi + R V' + U' \, ,\nonumber\\
\dot\phi^2 &\ll& 2U  \, .\nonumber
\ee
Within this approximation we find
\be
\dot\phi_1 = \sqrt{\frac{8}{3}}\,m_{Pl} m_\phi K
\sqrt{-\phi_s \phi_1}\,e^{K \phi_1/2} \, .
\label{dot-phi1-slow}
\ee
One should check of course that the neglected terms are sub--dominant, which is
indeed the case.

Introducing new dimensionless function $z = K\phi_1$ and frequency:
\be
\omega_1 = \frac{16\sqrt{2}}{\sqrt{3}}\,\frac{m_\phi m_{Pl} \phi_s^5}{\mu^6} \, ,
\label{omega}
\ee
we arrive at a very simple equation:
\be
\dot z = \omega_1\,\sqrt{-z} e^{z/2} \, .
\label{dot-z}
\ee
The characteristic time of evolution defined as $\tau_1 = 1/\omega_1$ is
small even in microscopic scale to say nothing of the cosmological one.
Though one should remember about exponentially small factor, $\exp (z/2)$,
which makes effective time much larger, it still seems non trivial to make it longer then the age of the Universe. In fact, for large $z$, the evolution is indeed slow enough, but at the same time the effective Planck mass becomes extremely tiny, thereby rendering our description unreliable. Nevertheless it might still be possible to find a solution which looks like our universe in the close proximity of $\phi_c$. We comment on this possibility in the concluding section.

Let us now turn our attention to equation (\ref{ddot-phi}) at the stage when $V''\gg 1$
and $(V')^2 \gg V$, when the equation takes the form:
\be
\ddot \phi + 3H\dot \phi + \frac{V''}{V'} \dot\phi^2 +
\frac{U'V - 2 UV'}{3(V')^2} = 0 \, .
\label{ddot-phi-2}
\ee
One can check that the term $\dot\phi^2\,V''/V'$ enforces very small value of
$\dot\phi$ and thus the Hubble parameter takes approximately the standard form, proportional
to square root of the energy density. One should of course
verify a posteriori that these terms can be neglected.
For $|\phi_1|<|\phi_c|$ the error we make is of order 20\%.

Neglecting $\ddot\phi$ and $(\dot\phi V'/2V)^2$ in $H$ we obtain the following
equation:
\be
\dot \phi = \frac{\sqrt{6}}{9}\,\frac{V^{3/2}}{(V')^2}\,
\frac{2KU - U'}{\sqrt{U}} \, .
\label{dot-phi2}
\ee
Now both terms in $U$ are essential, see eq.~(\ref{U-of-phi-1}), while the
factor $(2KU - U') $ takes the form:
\be
2KU - U' \approx -2K m_\phi^2 \phi_s \phi_1\,\left(1-\frac{1}{2K\phi_1}\right) \approx
-2K m_\phi^2 \phi_s \phi_1 \, .
\label{2KU}
\ee
Our approximation breaks down when $U\simeq0$, or, better to say, 
when $2K\phi_1 \simeq 1$, i.e.\ when
\be
\phi_1^{(0)} = - \frac{\mu^4 m_\phi^3}{16\sqrt{2} M_v^6} \, .
\label{phi1-0}
\ee
When $\phi_1$ approaches $\phi_1^{(0)}$, the $\dot \phi^2$--terms in $H$
should be taken into account.
The potential energy at $\phi_1^{(0)}$ is still positive but extremely tiny
\be
U_0 = \frac{\mu^8 m_\phi^8}{1024M_v^{12}} \, .
\label{U_0}
\ee

At some moment, very soon after $\phi_1$ has crossed $\phi_1^{(0)}$, the total energy density will vanish, $H$ will become zero, and
the cosmological expansion will stop and turn into contraction.
Our universe has not yet reached this stage and we will take $\phi_1$ far
from $\phi_1^{(0)}$ and neglect the constant term in the potential (\ref{U-of-phi-1}).

Introducing again $z=K\phi_1$ we find the equation very similar to eq.~(\ref{dot-z}):
\be
\dot z = \omega_2 \sqrt{- z} \,e^{-z/2} \, .
\label{dot-z2}
\ee
Here
\be
\omega_2 = \frac{1}{3\sqrt{3}}\,\frac{m_\phi^2 \mu^2}{m_{Pl} M_V^2} \, .
\label{omega-2}
\ee
Notice that the sign of the exponent is opposite to that in eq.~(\ref{dot-z}).

The characteristic time of evolution
\be
\tau_2 = \frac{1}{\omega_2} = 10^{21}\,{\rm s}\,\,\frac{m_\phi^2\, \mu^2}{\rm{TeV}^4}\,
\frac{(10^{16}\,{\rm TeV})^3}{m_{Pl}M_v^2} \, ,
\label{tau-2}
\ee
which is rather close to the cosmological time. The evolution of $z$
is much faster for very large negative $z$ but if $z\sim(-10) $ we remain with
long effective time of variation and still far from $\phi_1^{(0)}$, since
$z^{(0)} = K\phi_1^{(0)} = -1/2$.

The time variation of the Newtonian constant at this epoch is
\be
\frac{\dot G_N}{G_N} = - \dot z \, .
\label{dot-G}
\ee
Using $m_\phi = 3$TeV, according to eq.~(\ref{dot-z2}) it is smaller than 
$10^{-19}$/sec for $z=-10$ and
is perfectly below the existing bounds~\cite{gn-bound}. See also the 
recent paper~\cite{cosimo} for more stringent, although speculative, limits.

It is noteworthy that at the stage when $\phi_1$ is still far from $\phi_1^{(0)}$,
the vacuum-like energy is positive, noticeably larger
than that given by eq.~(\ref{U_0}), and rather naturally close to the
observed value.

The problem with this second evolutionary regime is that, despite the presence of vacuum 
energy of the correct magnitude, the Universe is not accelerating. It is easy to see how 
this comes about by looking again at equation (\ref{acceleration}), which in the regime of 
interest here reads $\ddot a/a \simeq - U/6V$, thereby implying a negative acceleration. 
One can also check that this must be the case by the expressions for energy and pressure 
densities (\ref{rho-phi}) and (\ref{p-phi}), which in this regime can be cast in the form
\be
\rho_\phi &\simeq& U - 6 H \dot\phi V' \, ,\nonumber\\
3 P_\phi &\simeq& U + 6 H \dot\phi V' \, .\nonumber
\ee
Using (\ref{dot-phi2}) (and accounting for the error made in that approximation), the equation of state therewith obtained confirms this result.

\section{Discussion/Conclusion}

We proposed here a model of dynamical adjustment of vacuum energy which quite
successfully achieves this goal, however at the high price of a too fast time variation of the gravitational coupling strength. Vacuum energy is compensated from initially huge
value down to some very small non--zero residual vacuum energy which magnitude is
naturally of the order of that observed today. The smallness of
remaining $\rv$ is dictated by its huge initial value - the surviving remnant
of $\rv$ is inversely proportional to its initial magnitude.
Of course there remain the problem of coincidence of vacuum energy today
and the energy density of the normal matter, which has not been considered here.

The asymptotic value of vacuum energy is constant and negative and it may seem to be
in contradiction with the data. However, the evolution of the compensating field,
which was very fast initially, drastically slows down at later cosmological stage
so that the universe remains with effective and slowly time--changing effective
vacuum energy during very long time which can be easily larger than the universe age
$t_U \approx 5\cdot 10^{17}$ sec. However in such a regime the Universe would be decelerating, in contrast to what we observe.

Going further, the model predicts that in a very distant future the universe will approach negative total
energy and the cosmological expansion will turn into contraction. This unexpected
feature differentiates our model from other adjustment scenarios but, unfortunately there is
not enough time in at our disposal to observe such a contraction.

For positive initial vacuum energy the potential of field $\phi$ is not
bounded from below, $U(\phi) = -m_\phi^2 \phi^2 + \rv$ but this feature
does not lead to any undesirable consequences because the equation of motion
has a stable minimum at finite $\phi = \phi_s$ where the energy density is finite
as well.

We have not considered realistic cosmology with the usual matter. Cosmology
with relativistic matter does not appear too tricky a problem, because
$T_\mu^\mu = 0$ and does not enter into expression for curvature (\ref{r-of-phi}). In the regime
when $(V')^2\gg V$, we obtain $R\ll H^2$, which is true for the usual relativistic
expansion, when $R=0$ and $H=1/2t$. However, if we have nonrelativistic matter
at the epoch when $(V')^2\gg V$, then again $R\ll H^2$ and the regime of cosmological
expansion would be determined again by $H\approx 1/2t$ which differs from the
usual one with $H= 2/3t$.

However at the moment a more urgent and serious problem is the too fast evolution of the effective Planck mass in the first regime. Possibly for realistic cosmology we need to study some other forms of potentials $U$ and $V$ which would allow to have a slow expansion regime when $(V')^2 \ll V$. On the other hand, this may be achieved by equipping the field with some decay channels or by studying the effects (enhancement of friction term) of gravitational particle production in that regime. If this could indeed be engineered, also matter would be readily and easily introduced.

{\bf Acknowledgments} We wish to thank C. Bambi for collaborating on the early stage of this work.

\end{document}